\documentclass[journal]{IEEEtran}
\usepackage[switch,columnwise]{lineno}
\usepackage{graphicx} 
\usepackage{cite}
\ifCLASSINFOpdf
\else
\fi
\usepackage{amsmath}
\usepackage[dvipsnames]{xcolor} 
\begin{document}
% \linenumbers
\title{Fully Convolutional Networks for Monocular Retinal Depth Estimation and Optic Disc-Cup Segmentation}

\author{ Sharath M Shankaranarayana, Keerthi Ram, Kaushik Mitra, Mohanasankar Sivaprakasam}

%Michael~Shell,~\IEEEmembership{Member,~IEEE,}
%        John~Doe,~\IEEEmembership{Fellow,~OSA,}
%        and~Jane~Doe,~\IEEEmembership{Life~Fellow,~IEEE}% <-this % stops a space
%\thanks{M. Shell was with the Department
%of Electrical and Computer Engineering, Georgia Institute of Technology, Atlanta,



\markboth{Journal Article}%
{Shell \MakeLowercase{\textit{et al.}}: Bare Demo of IEEEtran.cls for IEEE Journals}



%-- Title
%-1 ) Single deep architecture for multiple inverse problems 

\maketitle

\begin{abstract}
Glaucoma is a serious ocular disorder for which the screening and diagnosis are carried out by the examination of the optic nerve head (ONH). The color fundus image (CFI) is the most common modality used for ocular screening. In  CFI, the central region which is the optic disc and  the optic cup region within the disc are examined to determine one of the important cues for glaucoma diagnosis called the optic cup-to-disc ratio (CDR). CDR calculation requires accurate segmentation of optic disc and cup. Another important cue for glaucoma progression is the variation of depth in ONH region. In this work, we first propose a deep learning framework to estimate depth from a single fundus image. For the case of monocular retinal depth estimation, we are also plagued by the labelled data insufficiency. To overcome this problem we adopt the technique of pretraining the deep network where, instead of using a denoising autoencoder, we propose a new pretraining scheme called pseudo-depth reconstruction, which serves as a proxy task for retinal depth estimation. Empirically, we show pseudo-depth reconstruction to be a better proxy task than denoising. Our results outperform the existing techniques for depth estimation on the INSPIRE dataset.

To extend the use of depth map for optic disc and cup segmentation, we propose a novel fully convolutional guided network, where, along with the color fundus image the network uses the depth map as a guide. We propose a convolutional block called multimodal feature extraction block to extract and fuse the features of the color image and the guide image. We extensively evaluate the proposed segmentation scheme on three datasets- ORIGA, RIMONEr3 and DRISHTI-GS. The performance of the method is comparable and in many cases, outperforms the most recent state-of-the-art.  
\end{abstract}


% \fi
\IEEEpeerreviewmaketitle

\section{Introduction}
%\IEEEPARstart{T}{his} 

Glaucoma is a widely occurring eye disorder and poses a serious threat to vision. It is caused due to an increased intra-ocular pressure near the optic nerve head (ONH) region. The effect of pressure on ONH can be seen using various modalities, one of which is the color fundus imaging. The diagnosis of glaucoma using fundus imaging is based on the examination of ONH, which mainly involves delineation of optic disc-cup boundary and subsequent calculation of morphological information such as optic cup to disc ratio. 
Glaucoma progression is associated with the loss of optical fibres with a corresponding change in the optic disc (OD). Therefore, the region within the OD called the optic cup (OC) is subsequently enlarged which is called the phenomena of cupping. Hence, OC is characterized by depth \cite{Clinicalref1, Clinicalref2}. The depth provides an important cue for glaucoma detection as it enables us to visualize the cupping in the optic nerve head region which is the cause for the progression of the disease. But the color fundus image lacks depth information since it is just a 2D projection of a retinal surface.
%It has been studied that depth also characterizes the changes in optic nerve head and thus serves as an important cue for glaucoma detection . 
But the explicit measurement of the depth map is only possible through techniques such as stereo imaging or other imaging modalities such as optical coherence tomography (OCT). These techniques are currently not well-suited for large-scale screening on account of their higher costs and limited availability. Hence, depth estimation from a monocular color fundus image is necessary and is also a motivation for this work.

In summary, we propose a deep learning based framework for estimation of depth from a monocular fundus image. The only available retinal dataset for suitable depth estimation is INSPIRE-stereo \cite{PAMIstereoref}. It consists of $30$ retinal images with OCT-based ground truth depth. But the dataset lacks pixel-wise optic cup and disc annotations which is necessary for the segmentation task. Hence, we train the depth estimation network using INSPIRE-stereo dataset and then employ the trained network to predict depth for other datasets namely ORIGA \cite{origa}, RIMONEr3 \cite{rimone}, DRISHTI-GS \cite{drishti}, which contain pixel-wise optic disc-cup annotations.   %Talk about INSPIRE-stereo dataset- the only depth dataset, use for training, inference on other dataset%
\begin{figure}[tp]
\begin{minipage}{1.0\linewidth}
  \centering
  \centerline{\includegraphics[width=\linewidth]{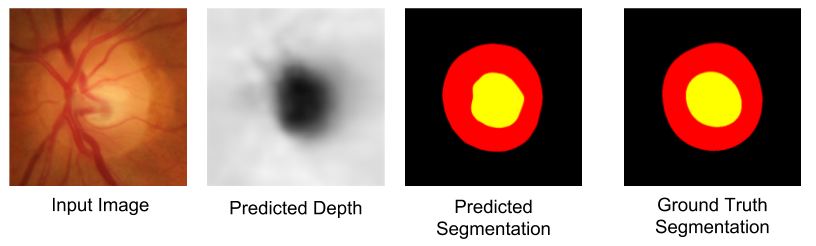}}
  %\vspace{-0.350cm}
% \centerline{(a) Result 1}\medskip
    \caption{Sample results from our method for an image from RIMONEr3 dataset.  Given an input RGB retinal image we first estimated the depth map using a deep network. Then we used the predicted depth map as the guide along with the RGB retinal image for semantic segmentation of optic disc and cup. }
%\vspace{-0.250cm}    
\end{minipage}
\label{Fig. Figure1_ex}
%\vspace{-0.40cm}
\end{figure}
We propose a novel guided network that uses the estimated depth as the guide in addition to the color fundus image to aid in optic disc and cup segmentation. Figure. 1 depicts the sample results for the prescribed work-flow.
% Glaucoma progression is associated with the loss of optical fibres with a corresponding change in the optic disc (OD). Therefore, the empty space within the OD and the so-called cup is subsequently enlarged which is a phenomenon called as cupping. That is the reason why the cup to disc ratio (CDR), defined as the relation between the OD and cup area, increases with the development of the disease. The depth estimation on retinal image provides an important cue for glaucoma detection as it enables us to visualize the intensity of the cupping region in optical nerve head region which is the reason for the progression of the disease. Hence depth estimation for retinal images becomes ultimately necessary and is also the motivation for this work. We aimed at estimating 3—D depth from a single retinal image as single image does not provide any depth cue itself. It is for the first time a  Fully Convolutional network is employed to learn the RGB image and its depth map features to predict depth from a single RGB fundus image.
In summary, the contributions of this work are as follows

\begin{enumerate}
    \item We propose a fully convolutional network for depth estimation from a single RGB fundus image and a novel block called Dilated Residual Inception (DRI) for simultaneous multiscale feature extraction. 
    \item We propose a simple pretraining scheme that provides improved results for depth estimation as compared to pretraining using denoising autoencoder which also addresses the problem of limited availability of ground truth depth data in the case of retinal imaging.
    \item We propose a fully convolutional guided network for the task of semantic segmentation of optic disc and cup. We also propose a multimodal feature extraction block which extracts and fuses image and depth information from two different modalities.
    \item We also explore the suitability of conditional random field (CRF) modelling on the image intensity in combination with depth, for refining the segmentation results.
    \item We extensively evaluated the proposed methods on three different datasets and have tabulated state of the art results.
    
\end{enumerate}


\label{sec:RELATED}
\section{Related Work}
For CDR estimation, a common pipeline employed is the detection of the optic disc (OD) region, followed by Optic disc segmentation and optic cup segmentation.

OD can be seen as a prominent circular region in the fundus images. It is generally brighter compared to the surrounding regions in a fundus image. Because of these characteristics, one of the most common techniques employed is template matching, exemplified in  \cite{temp-matching} where the Hough Transform is applied on the features extracted from the morphological operations to fit an ellipse or a circle.
As an improvement over the template matching based techniques, deformable methods such as Snakes \cite{snakes} and level-sets \cite{levelsets} apply energy minimization based on handcrafted features. The features are generally based on some form of gradient information and hence sensitive to abnormalities like peripapillary atrophy around the optic disc. Further, they are very sensitive to initialization. There have also been classification-based methods where handcrafted features are extracted from superpixels \cite{superpixel} to classify each superpixel  belonging to either OD or background classes. These methods do not tend to be robust as the handcrafted have some innate limitations.

The level-set method \cite{levelsets} also managed to solve for Optic cup segmentation with features based on pallor information. But again, they don't tend to be robust in cases that lacked unmarked changes in pallor between the disc and cup. Vessel kinks in the ONH region have been found to be informative \cite{Rbend} for the OC segmentation task. Such vessel bends or kinks are found using wavelet transform or curvature information,  and these approaches appear to address a difficult sub problem of accurate vessel bends and kinks detection in the context. Moreover,  the consistency of assumptions for the vessel bend to lie on the cup boundary might be data specific. 

Depth discontinuity of the retinal surface near the OC region is seen to be an important cue for glaucoma detection. The depth information is either obtained using modalities like OCT or stereo. 
There have been very few works presented on estimating depth from color fundus image \cite{PAMIstereoref,DepthDictionary,DepthAkshaya} compared to the numerous works presented for depth estimation in generic scenes using deep learning \cite{depth1, depth2, DepthHuber1}. The work presented in \cite{PAMIstereoref} proposed a method to calculate depth from stereo. In \cite{DepthDictionary}, the authors proposed a method for single image depth estimation where they estimate depth from color, shading and also using a coupled sparse dictionary-based supervision method. The individual depth estimates are then combined to give final depth map. In \cite{DepthAkshaya}, the authors proposed a combination of fast marching based depth estimation that relies on intensity image and supervised depth estimation from cup confidence map, where confidence map is obtained by coarse segmentation of cup.% None of these depth estimation techniques are robust and also are not end-to-end.  
  The obtained depth information has been used for Optic cup segmentation in \cite{DepthDictionary} using conditional random fields. Also recently, in \cite{depthmultimodal} authors proposed a method for OD-OC segmentation by using multimodal information. They used hand crafted features for training the classifier. 

Very recently, there have been several works based on deep learning for OD and OC segmentation. In \cite{zilly2017glaucoma}, the authors proposed a method where convolutional neural networks (CNNs) are used to learn filters in a greedy manner and then used them for feature extraction following which they obtained pixelwise predictions and final segmentation map using graph cut and convex hull transformations. The network contains fully connected layers and they are also not end-to-end, because of the pipeline of different steps involved. In \cite{Resunet}, the authors proposed a fully convolutional end-to-end OD-OC segmentation method. Recently, in \cite{PT2018} the authors proposed a multiscale network based on a modified U-net \cite{unet} architecture for OD-OC segmentation. They employed polar transformation (PT) on the RGB fundus images and segmentation map, before feeding the images to the network and they finally used inverse PT on the output. State of the art results were reported using PT but the network is not end-to-end. We  proposed a framework to first estimate depth from a single retinal image and then used an end-to-end network to perform multimodal fusion of features, i.e., combining depth and color image features. %To the best of our knowledge, such a framework has not been proposed for OD-OC segmentation. 

\begin{figure}[t!]
   \centering
\includegraphics[scale=0.25]{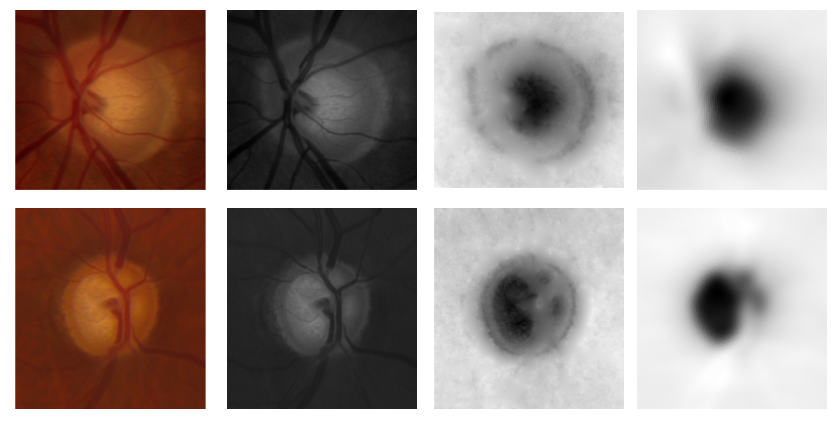} 
  \caption{The first column shows RGB fundus images, the second column shows the corresponding green channels, the third column shows the proposed pseudo-depth image and the last column shows the ground truth depth image. It can be seen that the pseudo-depth image looks very similar to the ground truth depth map}
  \label{Fig. PseudoDepth}
\end{figure}
%-----------------------Figure----------------------------------

%--------------- End Figure
\section{Methods}
%%%%%%%%%% Depth Subsection
%----------- Figure 
\begin{figure*}[htb!]
   \centering
\includegraphics[width=\linewidth]{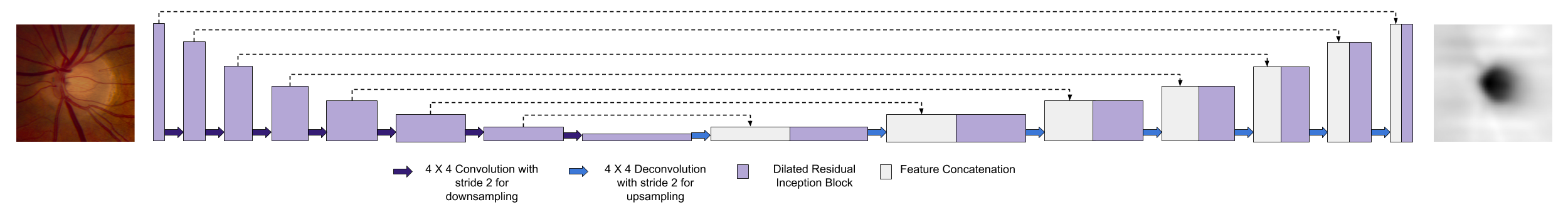} 
  \caption{Depth Estimation Network: Encoder decoder architecture with special blocks for extracting multiscale information simultaneously at each level. The network takes in RGB fundus image as input and predicts depth map as output. }
  \label{Fig. Depth Estimation}
\end{figure*}
%-----------------Figure
\begin{figure}[htb!]
   \centering
\includegraphics[scale=0.25]{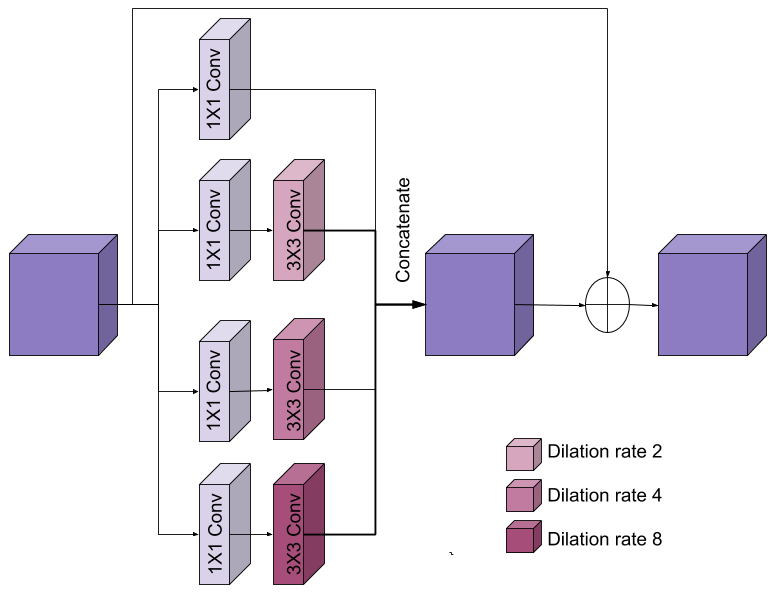} 
  \caption{Dilated Residual Inception Block}
  \label{Fig. DRIB}
\end{figure}
%------------
%--- End Figure
\subsection{Depth estimation}
The first important task in the proposed pipeline is monocular retinal depth estimation. We proposed a fully convolutional end-to-end network for depth estimation task. 
As such,  single image depth estimation is a challenging task and unavailability of a large depth dataset for retinal images makes it even more challenging. Denoising autoencoders \cite{Autoencoder1} have been commonly used as a method for unsupervised pretraining, where the image representations are learned by reconstructing clean images from its corrupted counterparts i.e., given a clean image $I$, it is corrupted by noise giving $I_\eta$,  and the representation is learned by the network while minimizing the following $L_2$ reconstruction loss-
\begin{equation}
\label{eq1}
L_{recon loss}=\|F(I_\eta) - I\|_2
\end{equation}

where $F(.)$ is the deep network which is to be pretrained. 
Thus, looking in an alternate way, denoising is used as a proxy task for learning representations. Very recently, there have been lot of works that has explored other ways of self-supervision for learning representations \cite{ContextAE, Colorization}.\\
For retinal image depth estimation, we explored the possibility of a better way of self-supervision. Upon close examination, we found that for a color fundus image $I_{RGB}$ with each of the channels individually normalized, the inverted green channel i.e., $(1 - I_G)$, (where $I_G$ is the green channel) with the vessels in-painted closely resembles to the depth map Figure. \ref{Fig. PseudoDepth}. We also experimented with the red and blue channels by performing the same kind of transformations as done for the green channel and found the transformed green channel to be closest to the depth map. We call this transformed green channel image as pseudo-depth image $I_{PD}$. We learned the representations while minimizing the following reconstruction loss
\begin{equation}
\label{eq2}
L_{recon loss}=\|F(I_{RGB}) - I_{PD}\|_2
\end{equation}
In other words, we learned the representations by reconstructing pseudo-depth image from a color fundus image. This is intended to be a better weight initialization for depth estimation compared to using denoising to learn the representation. %This serves as a better weight initialization scheme for depth estimation task than using a denoising autoencoder. 

\subsubsection{Network architecture}
The architecture for the fully convolutional network for depth estimation is shown in Fig. \ref{Fig. Depth Estimation}. The network structure is similar to standard encoder-decoder architectures employed for various tasks. The network consists of special blocks which are inspired from \cite{resnet} and \cite{inception} , the details of which are discussed in the subsequent sub-section.
 For encoder part of the network, we used $4 \times 4$ convolution with a stride of $2$ followed by batch normalization, leaky ReLU (with slope $0.2$) operations for downsampling and double the number of filters after downsampling, but doubling is done only till the number of filters reach $512$. This allows us to have a deeper network with relatively less parameters. Also strided convolution is employed instead of regular max-pooling operations to aid in smoother gradient flow. For decoder part of the network, we used feature concatenation operations similar to the original U-net and again used both convolutional and residual blocks. For the first $3$ decoding layers, we used dropout with drop rate of $50\%$. We did a $4 \times 4$ deconvolution operation with stride $2$ for upsampling, followed by batch norm and ReLU operations. After the last decoding layer, a $1 \times 1$ convolution is performed to the number of output channels of the map followed by $tanh$ activation.

%%%%%%%%%%%%%%
\begin{figure*}[htb!]
   \centering
\includegraphics[width=\linewidth]{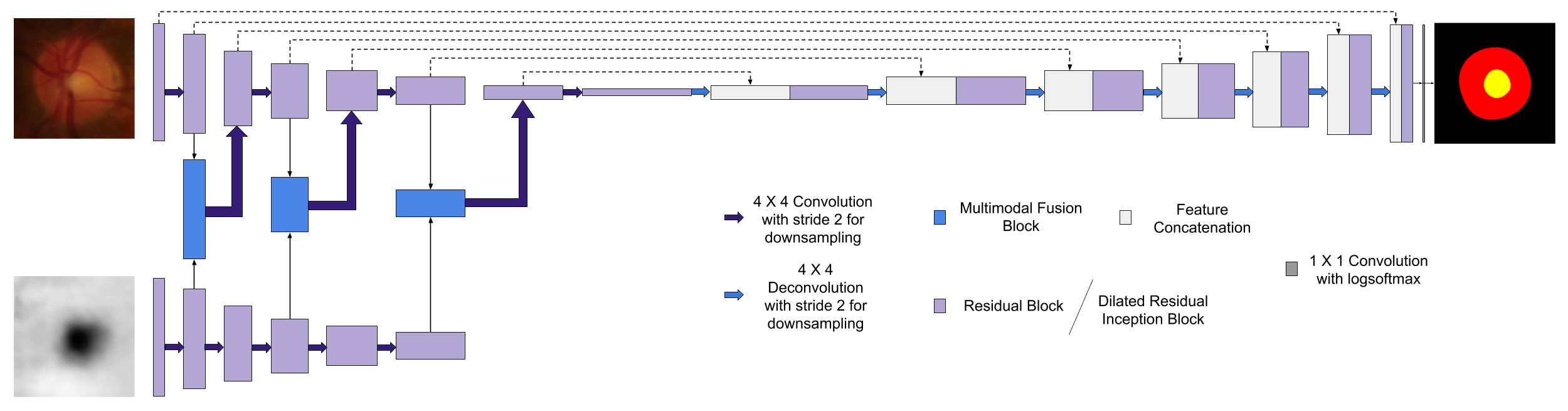} 
  \centering
  \caption{Deep guided semantic segmentation network which is used to perform OD and OC segmentation using RGB image and with another image such as a depth map as a guide.}
  \label{Fig. Depthguidedsegment}
\end{figure*}
%%%%%%%%%%%%%%%%%
%-------------------------------------
\begin{figure}
   \centering
\includegraphics[width=\linewidth]{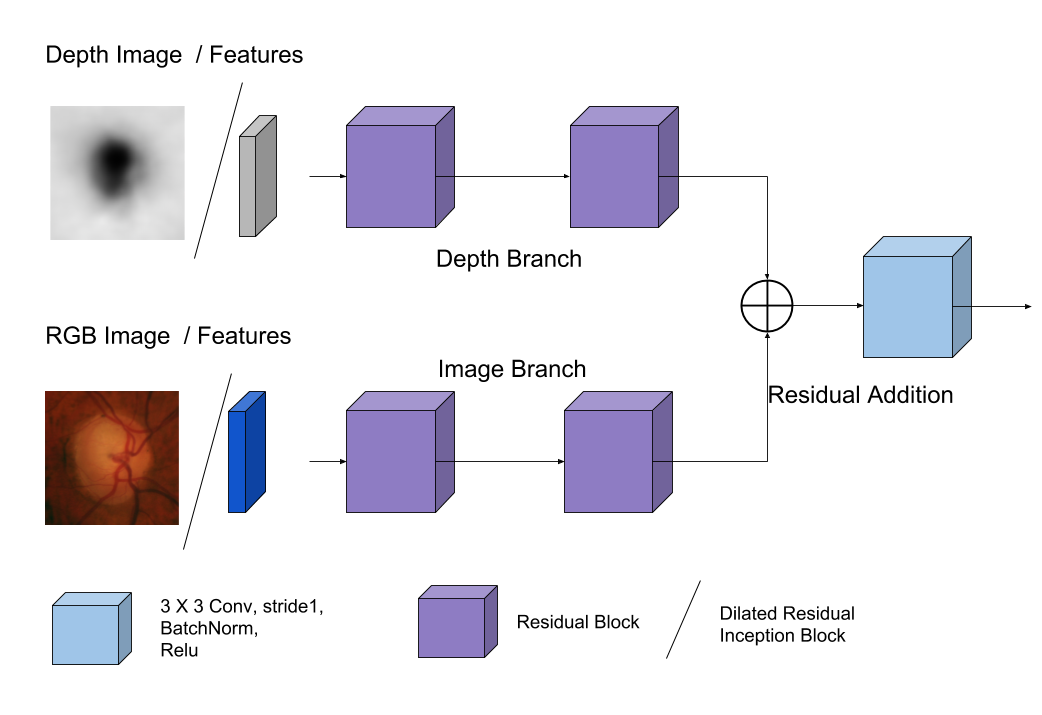} 
  \centering
  \caption{Depth and image feature extraction block of the depth guided fully convolutional network }
  \label{Fig. DGN}
\end{figure}
%%%%%%%%%%%%%%%%%
\subsubsection{Dilated residual inception blocks}
The special blocks employed in the main architecture are inspired from inception module \cite{inception} and dilated convolution \cite{DilatedMain}. The inception module proposed in GoogleNet architecture is a method to fuse multi-scale information. The module consists of multiple convolution operations with various kernel size performed in parallel and the result of all convolutions is then concatenated. We propose a modification for this inception module by incorporating dilated convolutions which are known to increase the receptive field exponentially with linear increase in the number of parameters \cite{DilatedMain}. Deep networks having dilated convolutions have been very recently employed in semantic segmentation \cite{Atrous1}, \cite{Atrous2} reporting an improved performance. Shown in Fig. \ref{Fig. DRIB}, the proposed inception module has an advantage that it has fewer parameters than the original inception module. Lastly, we also incorporate residual connections \cite{resnet} since they alleviate the problem of vanishing gradients in deep networks and also help in faster convergence.  

\subsubsection{Loss functions}
As our first task, we train a fully convolutional network $F$ to predict depth from a single RGB fundus image. We solve for depth estimation as a regression problem. It should be noted that the network at this stage is already pretrained using the scheme discussed in the earlier subsection III.A. Here, given an RGB fundus image $I$ and the corresponding ground truth depth map $d$, our network learns the mapping $F_{depth}: I \rightarrow d$. As in the case of many regression formulations, we employed the standard $L_2$ loss function. 
\begin{equation}
    L_2(\Hat{d},d) = \|\Hat{d} - d\|_2
\end{equation}
where $\Hat{d}$ is the network output $F_{depth}(I)$. 
But since it has been reported that reverse Huber loss (berHu) \cite{BerHu1} results in improved depth estimation in \cite{DepthHuber1}, we also experimented with berHu loss $L_{berHu}$ and also $L_1$ loss ($L_1(\Hat{d},d) = \|\Hat{d} - d\|_1$) for completeness. BerHu loss is given by:
\begin{equation}
    L_{berHu}(\Hat{d},d)=\begin{cases}
    |\Hat{d} - d| & |\Hat{d} - d| \leq c, \\
    \frac{{(\Hat{d}-d)}^2 + c^2}{2c} & |\Hat{d} - d| > c
    \end{cases}
\end{equation}
%%%%%%%%%%%%%%

where $c$ is set to $\frac{1}{5}\max_i(|\Hat{d_i} - d_i|)$, with $i$ indexing all the images in a batch, i.e., $c$ set to $20\%$ of the maximal per-batch error, similar to \cite{DepthHuber1} .
%%%%%%%%% End Depth Subsection
%%%%%%%%%%%%% Start Segmentation Section
\subsection{Optic disc and cup segmentation}
The second major task in our pipeline is to perform semantic segmentation on optic disc and cup for a given retinal image. In addition to performing semantic segmentation using image features, we propose a scheme to extract depth features along with the image features in a fully convolutional framework setting. The proposed scheme can be employed for feature extraction and fusion from any two modalities and in our case for OD-OC segmentation, we use depth map as the guide image along with RGB fundus image.

\subsubsection{Multimodal feature fusion block}
The schematic of multimodal feature fusion block is shown in figure \ref{Fig. DGN}. Although, it can handle images of any two different modalities, in the presented work we mainly use RGB image and the depth map. Thus, the network consists of two branches- the depth branch and the image branch. The input to the image branch is either an image or its features and is the same for depth branch.  
Each of the branches employ special blocks for feature extraction. We experimented with two kinds of special blocks - the simple residual block \cite{resnet, Resunet} and dilated residual inception (DRI) block (Figure. \ref{Fig. DRIB}). 
We used two successive special blocks in both the branches and then performed residual pooling where the output of depth branch is added element-wise with the output of image branch. The resultant output is then passed to a $3 \times 3$ Conv-Batchnorm-ReLU block. 

\subsubsection{Depth Guided Semantic Segmentation Network}

The base architecture for the task of semantic segmentation is similar to the fully convolutional network proposed in \cite{Resunet}, with a difference that we also experimented with the DRI blocks in place of simple residual blocks. Also, the previous network only takes in color fundus image as input, whereas we propose a scheme to extract both depth features and image features and then fuse them using the multimodal feature extraction block. The schematic of the segmentation network is shown in Fig. \ref{Fig. Depthguidedsegment}. The network consists of a main branch with an encoder-decoder architecture similar to \cite{Resunet} but consists of an additional encoder part to incorporate the additional depth input. The depth encoder has same structure as the main branch except it consists of six levels as compared to eight levels in main branch. The output features of every alternate level from both the branches is passed through multimodal feature fusion block and the fused output is fed into the next level by using a $4 \times 4$ Conv with stride 2 for downsampling and also the number of filters is doubled. It should be noted that the passing of fused information is done only in the main branch and not in the depth branch. We did not perform fusion at every level following \cite{Fusenet} where it was shown that the sparse fusion gave better results than dense fusion. Also, only the features from main branch are passed to the decoder while using long skip connections for feature concatenation from the encoder part of the network to the decoder part. The network is trained with multiclass cross-entropy loss given by-
\begin{equation}\label{segloss}
L_{mce}=-\sum_{c}^{C}{\sum_{i}^{N}y_{i}\log{(x_i)}}
\end{equation}
where $c$ represents the class index with total of $C$ classes, $i$ represents the pixel index with total of $N$ pixels and $y_{i}$ represents ground truth label map and $x_i$ represents probability map predicted by the network.

%------------------------------------

\subsubsection{CRF based Post-processing}
We also evaluated the effectiveness of CRF for the post-processing of network predictions. For this case, we do not consider the depth guided network instead we consider the network only with
RGB image as the input. With $I$ as the input image having a size $N$ and $\textbf{x}$ as the label vector, the Gibbs energy in a fully connected pair-wise CRF model is given by

%------------------------------------
\begin{equation}\label{CRF}
E(\textbf{x})=\sum_i\phi_u(x_i) + \sum_{i<j}\phi_p(x_i)
\end{equation}
%-----------------------------------

where $i$ and $j$ range from $0$ to $N$ and $\phi_u(x_i)$ is the unary potential which is the output of the network or in other words the negative log of probabilities for all the classes. The FCN forms the unary classifier in our case and the output for each pixel is independent from others.  
The pairwise potentials $\phi_p(x_i)$ have the form
\begin{equation}\label{Pairwise}
\phi_p(x_i)=\mu(x_i,x_j)k(f_i,f_j)
\end{equation}
%-----------------------------------
where $\mu$ is the label compatibility function given by Pott's model \cite{Potts} $\mu(x_i,x_j) = [x_i \neq x_j]$ and $k$ is a Gaussian kernel with $f_i$ and $f_j$ being feature vectors for pixels $i$ and $j$ respectively. Similar to \cite{CRF}, we use two kernel potentials. But, in addition to contrast sensitive potentials, we also use depth sensitive potentials. 

%------------------------------------
\begin{eqnarray}\label{Pairwise}
k(f_i,f_j) = w_1 \exp{(-\frac{|p_i-p_j|^2}{2\theta_{\alpha}^2}-\frac{|I_i-I_j|^2}{2\theta_{\beta}^2})} +  \nonumber \\
w_2 \exp{(-\frac{|p_i-p_j|^2}{2\theta_{\alpha}^2}-\frac{|d_i-d_j|^2}{2\theta_{\gamma}^2})} + \nonumber \\
w_3 \exp{(-\frac{|p_i-p_j|^2}{2\theta_{\alpha}^2})}
\end{eqnarray}
%-----------------------------------

where $p_i, p_j$ denote positions and $I_i, I_j$ denote image intensity vectors and  $d_i, d_j$ denote depth intensity vectors and $w_1, w_2, w_3 $ are weight terms. %The readers are advised to refer \cite{CRF} for further details on inference.

\section{Experiments and results}
\begin{figure*}[htb!]
   \centering
\includegraphics[width=\linewidth]{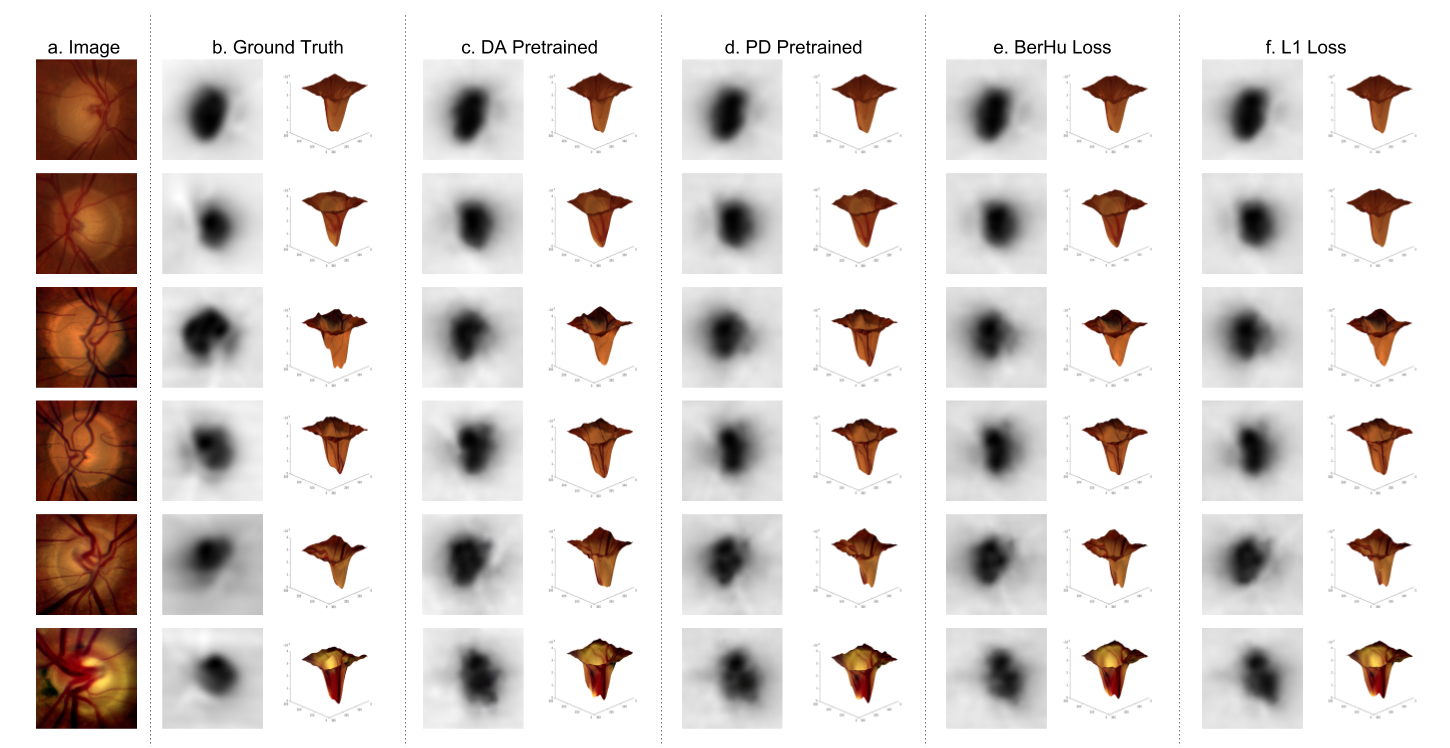} 
  \caption{Sample results of Depth Estimation }
  \label{Fig. DepthResults}
\end{figure*}

For the task of depth estimation, we first pretrained the network. For this we collected retinal images from multiple datasets such as INSPIRE-stereo \cite{PAMIstereoref}, ORIGA\cite{origa}, RIMONE \cite{rimone}, DRISHTI \cite{drishti}, DRIONS \cite{drions} and cropped the region of interest which in our case is the optic nerve-head region. We then standardized all the images individually by subtracting the mean and dividing by standard deviation for each of the color channels. To help alleviate the dataset bias, we enforce the mean and standard deviation of each image to be the one obtained from a canonical image. This ensures that intensity levels of images of various datasets roughly follow the same desired distribution. 
For comparing pretraining tasks, we experimented with a denoising autoencoder, for which we first created the corrupted and clean image pairs by adding uncorrelated white Gaussian noise to the images collected. We then considered the FCN to be used for depth estimation and trained it to reconstruct the clean image from the corrupted image. For the second case of pretraining, we trained the FCN to reconstruct pseudo-depth image from the color fundus image. Later, while estimating for depth, weights are initialized with the weights learned from either of the pretraining tasks instead of random initialization.
For depth estimation, we used the INSPIRE-stereo dataset \cite{PAMIstereoref}. It consists of 30 retinal images along with their depths. This is the only publicly available depth dataset for retinal images. We did a five-fold cross validation for this dataset. For the train split we do heavy data augmentation by using flip, zoom, noise jitter and in turn increased the number of images in the train split by ten-fold. We used the network shown in Figure.~\ref{Fig. Depth Estimation} for training and experimented it using all the three loss functions discussed. The results obtained are shown in Fig.~\ref{Fig. DepthResults} for qualitative evaluation

For quantitative evaluation, we use the root mean squared error (RMSE) given by

$RMSE =  \sqrt{\Sigma(x_i - y_i)^2}$

% \begin{equation}
%   RMSE =  \sqrt{\Sigma(x_i - y_i)^2}
% \end{equation}
and also and  correlation coefficient (Corr) $r$ given by 

$r(x,y) = \frac{\Sigma(x_i - \bar{x})(y_i - \bar{y})}{\sqrt{\Sigma(x_i - \bar{x})^2\Sigma(y_i - \bar{y})^2}}$

% \begin{equation}
%      r(x,y) = \frac{\Sigma(x_i - \bar{x})(y_i - \bar{y})}{\sqrt{\Sigma(x_i - \bar{x})^2\Sigma(y_i - \bar{y})^2}}
% \end{equation}

where $x$ and $y$ are estimated depth maps and ground truth depth maps respectively, and $i$ is the pixel index. The obtained values are tabulated in Table.I for four cases - 
\begin{itemize}
    \item Denoising autoencoder (DA) pretrained network with $L_2$ loss function,referred to as DA
    \item Pseudo-depth (PD) pretrained network with $L_2$ loss function, referred to as PD
    \item Pseudo-depth (PD) pretrained network with reverse Huber ($berHu$) loss function, referred to as berHu Loss
    \item Pseudo-depth (PD) pretrained network with $L_1$ loss function, referred to as $L_1$ Loss
\end{itemize}

%-------------------- Table 1 Depth
%---------------------
\begin{table}
\caption{Comparison of various Depth estimation methods }
  \centering
  \begin{tabular}{|c|c|c|c|c|}
    \hline
    {\textbf{Method}} & \multicolumn{2}{c|}{\textbf{Corr}} & \multicolumn{2}{c|}{\textbf{RMSE}}\\
    % \hline
    % \textbf{Inactive Modes} & \textbf{Description}\\
    \cline{2-5}
    & Mean & Std-Dev & Mean & Std-Dev\\
    %\hhline{~--}
    \hline
     Multi-scale stereo \cite{PAMIstereoref} & - & - & 0.1592 & 0.0879 \\ \hline
     Dictionary-based \cite{DepthDictionary} & 0.8000 & 0.1200 & - & - \\ \hline
     Fast marching based \cite{DepthAkshaya} &  0.8225 & - & 0.1532 & 0.1206 \\ \hline \hline
     DA with L2 loss & 0.9566 & 0.0265 & 0.0064 & 0.0043 \\ \hline
     PD with L2 loss & \textbf{0.9629} & 0.0222 & \textbf{0.0059} & 0.0030 \\ \hline
     PD with berHu Loss & 0.9595 & 0.0232 & 0.0060 & 0.0020 \\ \hline
     PD with L1 Loss & 0.9595 & 0.0241 & 0.0060 & 0.0032\\ \hline
 \end{tabular}
 %   \vspace{-2em}
  \end{table}
%-----------------

%%%%%%%%%%%%%%%%%%%%%%%%%%%%%%%%%%%%%%%%%%%%%%%%%%%%%%%%%%%%%%%%%%%%%
%%%%%%%%%%%% Figure Origa
\begin{figure*}
   \centering
\includegraphics[width=\linewidth]{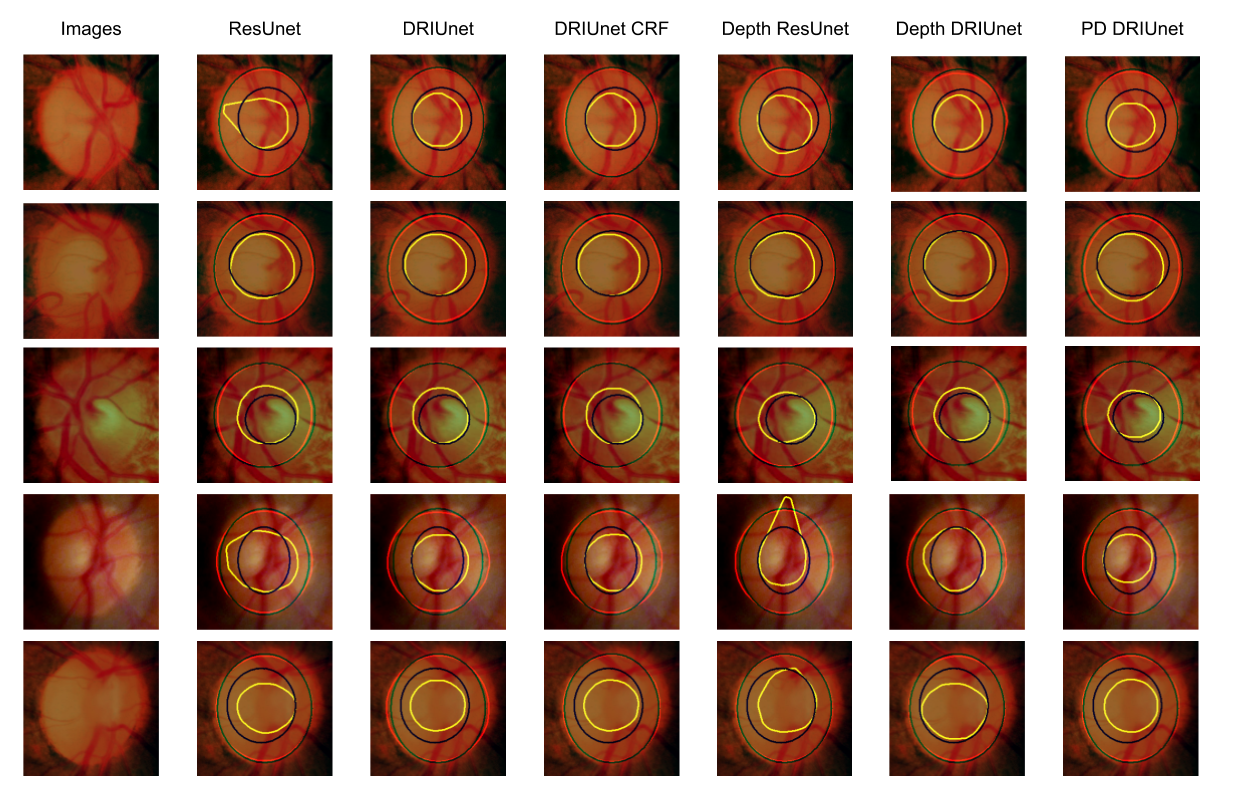}  \caption{Qualitative results for ORIGA dataset, along with the ground truth. The green and black lines indicate the ground truth segmentation boundaries for optic disc and cup respectively and red and yellow lines indicate the segmentation boundaries for optic disc and cup respectively obtained using various methods }
  \label{Fig. Origa}
\end{figure*}
%%%%%%%%%%%%%%%%%%%%%%%%%%%%%%%%%%

%%%%%%%%%%%%%%%%%%%%%%%%%%%%%%%%%%%%%%%%%%%%%%%%%%%%%%%%%%%%%%%%

%------------------Table 2 Origa
\begin{table*}
\caption{Comparison of various segmentation methods on ORIGA dataset }
  \centering
  \begin{tabular}{|c|c|c|c|c|c|c|c|}
    \hline
    {\textbf{Method}} & \multicolumn{3}{c|}{\textbf{Disc}} & \multicolumn{3}{c|}{\textbf{Cup}} &
    {\textbf{$\delta_E$}} \\
    \hline
    % \textbf{Inactive Modes} & \textbf{Description}\\
    \cline{2-7}
    & E & A & D & E & A & D &\\
    %\hhline{~--}
    \hline
     R-bend \cite{Rbend} & 0.129 & - & - & 0.395 & - & - & 0.154 \\ \hline
     Super-pixel \cite{superpixel} & 0.102 & 0.964 & - & 0.264 & 0.918 & - & 0.077 \\ \hline
     Unet \cite{unet} & 0.115 & 0.959 & - & 0.287 & 0.901 & - & 0.102 \\ \hline
     MNet + PT \cite{PT2018} & 0.071 & \textbf{0.983} & - & 0.230 & 0.930 & - & 0.071 \\ \hline \hline
     ResUnet & 0.047 & 0.978 & 0.976 & 0.232 & 0.921 & 0.863 & 0.069 \\ \hline
     DRIUnet & 0.047 & 0.977 & 0.976 & 0.219 & 0.927 & 0.873 & 0.068 \\ \hline
     DRIUnet CRF & \textbf{0.044} & 0.980 & \textbf{0.977}  & 0.219 & 0.927 & 0.873 & 0.068 \\ \hline
     Depth ResUnet & 0.047 & 0.977 & 0.974 & 0.221 & 0.926 & 0.870 & 0.071\\  \hline
     Depth DRIUnet & 0.051 & 0.975 & 0.974 & 0.216 & \textbf{0.930} & 0.874 & 0.069 \\ \hline
     PD DRIUnet & 0.051 & 0.974 & 0.972 & \textbf{0.212} & 0.928 & \textbf{0.876} & \textbf{0.067} \\ \hline
     
 \end{tabular}
 %   \vspace{-2em}
  \end{table*}
%-----------------
%%%%%%%%%%%% Figure Origa

From the Table.I, it can be seen that we achieve significant improvement for depth estimation over the previously proposed methods. Also, the pseudo-depth pretraining method gave much better results when compared to denoising autoencoder based pretraining method. This shows that pseudo-depth reconstruction from color fundus image is a better proxy task for depth estimation. But, contrary to reports performing depth estimation in natural images, the use of reverse Huber(berHu) loss did not improve the depth estimation results in our case. Also depth estimation with $L_1$ loss function seemed to give similar results as the berHu loss function. The reason for getting similar results with different loss function could be attributed to smoother variations in the depth maps of retinal images unlike the depth maps of natural images. 

\begin{figure*}[htb!]
   \centering
\includegraphics[width=\linewidth]{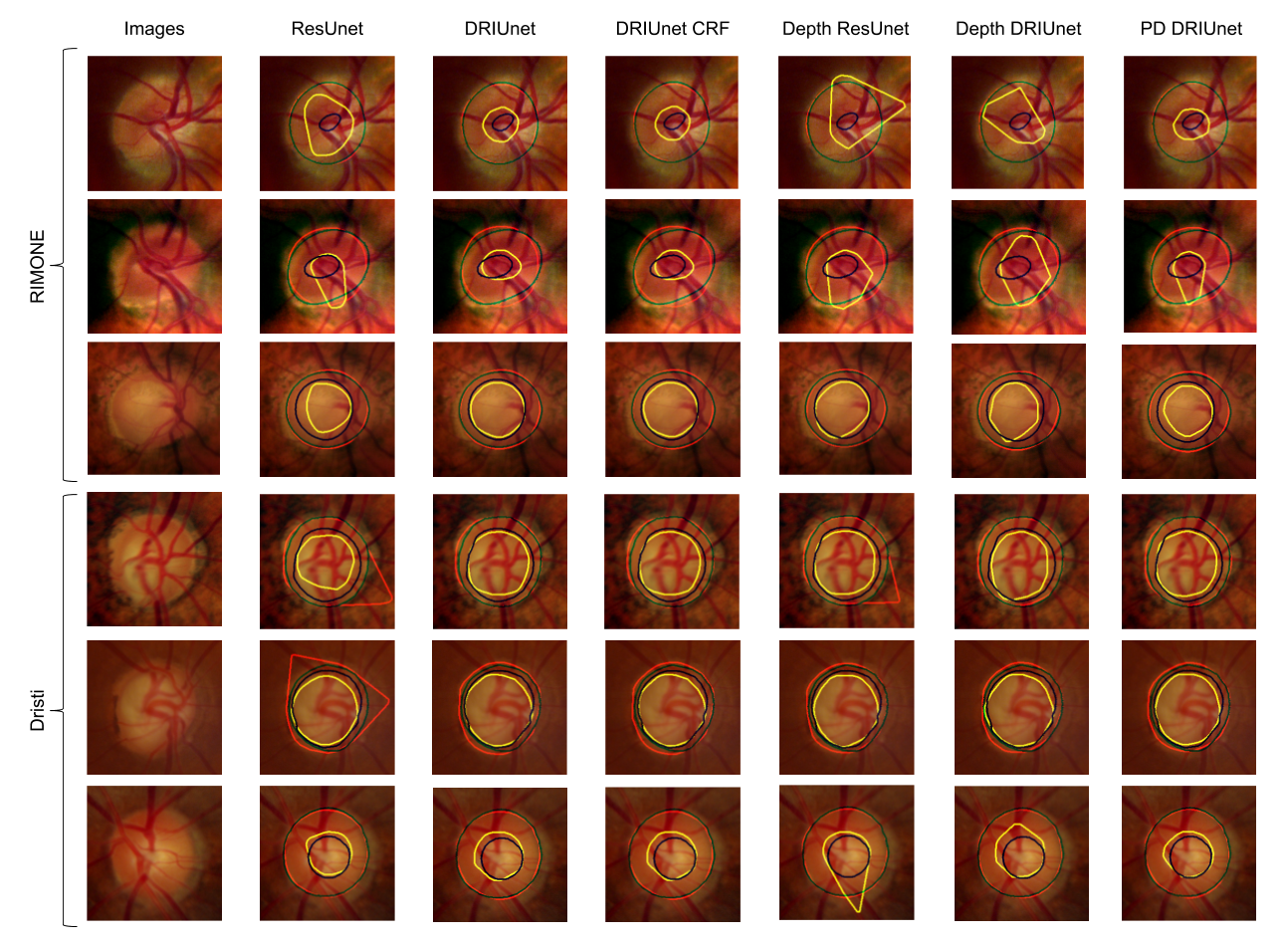}   \caption{Qualitative results for RIMONE and Drishti Datasets, along with the ground truth. The green and black lines indicate the ground truth segmentation boundaries for optic disc and cup respectively and red and yellow lines indicate the segmentation boundaries for optic disc and cup respectively obtained using various methods }
  \label{Fig. Rimone}
\end{figure*}

For the task of optic disc and cup segmentation, we used three datasets ORIGA, RIMONEr3 and DRISHTI. The first one contains $650$ retinal images along with the pixelwise markings of optic disc and cup. Similar to the work presented in \cite{PT2018} and for the purpose of comparison, we split the dataset with $325$ images for testing and the remaining for training. For training, we used standard data augmentation techniques to increase the number of images. We trained the network from scratch and we obtained the final segmentation map by thresholding the output probabilities similar to the work done in \cite{PT2018}. We then applied a convex hull transformation on the segmentation outputs for both cup and disc. The sample outputs for ORIGA dataset are shown in Figure.~\ref{Fig. Origa} where we show the delineations of the predicted and ground truth optic disc and cup segmentations, tabulated for different experiments. For RIMONE and DRISHTI datasets, we again divided the dataset into two halves, one for training and the other for testing. But given that RIMONEr3 and DRISHTI are smaller datasets, we performed fine-tuning instead of training from scratch, initializing the network with the weights obtained by training for the ORIGA dataset. It should be noted that the guided network requires an additional input- the depth map as guide image. Since none of these three datasets have the depth information, we obtained the depth maps for all the images of these datasets by passing the RGB fundus images through the depth estimation network trained on INSPIRE-stereo dataset. The sample outputs for these datasets are shown in Figure.~\ref{Fig. Rimone}.  
%We tried to alleviate the effect of dataset bias by having a canonical image and standardizing the color channels of all the different dataset with respect to this canonical image (hence the color differs when compared with the original dataset in all the figures with retinal images).     

We used the standard segmentation metrics such as overlapping error ($E$ (Eqn.9)) where S and G denote the segmented mask and the manual ground truth, balanced accuracy ($A$ (Eqn.10)), where $Sen$ (Eqn.11) and $Spe$ (Eqn.12) are the sensitivity and specificity. $TP$ and $TN$ denote the number of true positives and true negatives respectively, and $FP$ and $FN$ denote the number of false positives and false negatives and dice coefficient ($D$) for quantitative evaluation of segmentation outputs. Since for glaucoma detection, one of the important indicators is the vertical cup to disc ratio (CDR), we also calculated the CDR for our obtained results and computed the absolute CDR error ($\delta_E$) given by $|CDR_{GT} - CDR_O|$ where $CDR_{GT}$ and $CDR_O$ are the ground truth and output cup to disc ratios respectively.

%  $E = 1 - {\frac{Area(S \cup G)}{Area(S \cap G)}$
 
%  $A = \frac{1}{2}(Sen + Spe)$
%  \\ $Sen = \frac{TP}{TP+FN}$  \hspace{1cm}          $Spe = \frac{TN}{TN+FP}$

\begin{equation}
  E = 1 - \frac{Area(S \cup G)}{Area(S \cap G)}
\end{equation}
\begin{equation}
  A = \frac{1}{2}(Sen + Spe)
\end{equation}
\begin{equation}
  Sen = \frac{TP}{TP+FN}
\end{equation}
\begin{equation}
  Spe = \frac{TN}{TN+FP}
\end{equation}

% \vspace{2mm}

We reported the obtained results for the following experiments-
\begin{itemize}
    \item Network with residual blocks and no depth (ResUnet)
    %\item Network with residual blocks with CRF post processing (ResUnet CRF)
    \item Network with dilated residual inception  blocks and no depth (DRIUnet).
    \item Network with dilated residual inception  blocks with CRF post processing (DRIUnet CRF).
    \item Depth based network with residual blocks (Depth ResUnet)
    \item Depth based network with dilated residual inception blocks (Depth DRIUnet)
    \item Pseudo-depth based network with dilated residual inception blocks (PD DRIUnet)
\end{itemize}
\begin{figure*}[htb!]
   \centering
\includegraphics[width=\linewidth]{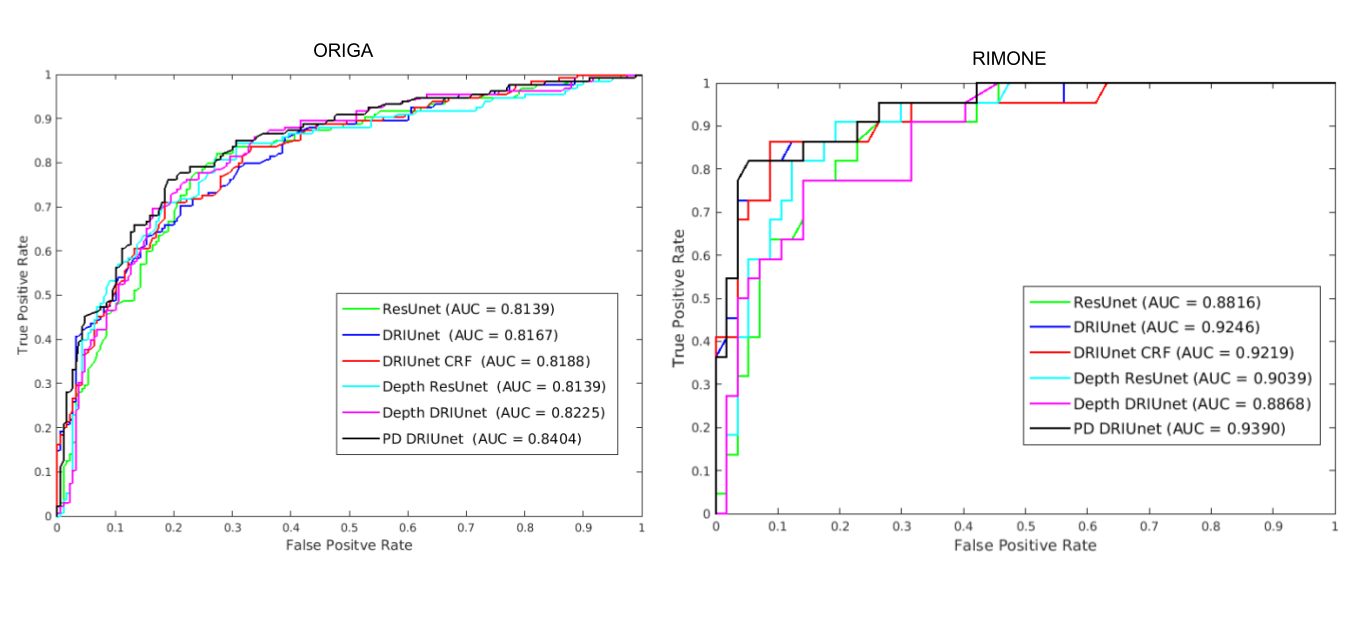} 
  \caption{Area under ROC Curves for ORIGA and RIMONEr3 datasets }
  \label{Fig. ROC}
\end{figure*}

%%%%%%%%%%%%%%%%%%%%%%%%%%%%%%%%%%
%
%------------------Table 3 RIMONE and DRISHTI
\begin{table*}
\caption{Performance of proposed segmentation method on RIMONE and DRISHTI datasets }
  \centering
  \begin{tabular}{|c|c|c|c|c|c|c|c|}
    \hline
    {\textbf{Method}} & \multicolumn{3}{c|}{\textbf{Disc}} & \multicolumn{3}{c|}{\textbf{Cup}} &
    {\textbf{$\delta_E$}} \\
    \hline
    % \textbf{Inactive Modes} & \textbf{Description}\\
    \cline{2-7}
    & E & A & D & E & A & D &\\
    %\hhline{~--}
    \hline \hline
     RIMONE & & & & & & & \\ \hline
     ResUnet & 0.060 & 0.974 & 0.968 & 0.321 & 0.914 & 0.863 & 0.086 \\ \hline
     DRIUnet & 0.060 & 0.974 & 0.968 & 0.284 & 0.925 & 0.873 & \textbf{0.065} \\ \hline
     DRIUnet CRF & 0.061 & 0.974 & 0.968 & 0.285 & 0.924 & 0.873 & 0.066 \\ \hline
     Depth ResUnet & 0.059 & 0.974 & 0.960 & 0.299 & \textbf{0.9353} & 0.816 & 0.081\\  \hline
     Depth DRIUnet & 0.059 & 0.975 & 0.969 & 0.310 & 0.910 & 0.874 & 0.082 \\ \hline
     PD DRIUnet &\textbf{0.058} & \textbf{0.975} & \textbf{0.970} & \textbf{0.284} & 0.920 & \textbf{0.876} & 0.066 \\ \hline
     DRISHTI & & & & & & & \\ \hline
     ResUnet & 0.089 & 0.968 & 0.952 & 0.283 & 0.926 & 0.813 &  0.118\\ \hline
     DRIUnet & 0.074 & 0.972 & 0.961 & 0.286 & 0.932 & 0.811 &  0.146\\ \hline
     DRIUnet CRF & 0.077 & 0.971 & 0.960 & 0.270 & 0.941 & 0.829 & 0.132 \\ \hline
     Depth ResUnet & 0.073 & 0.961 & 0.974 & 0.268 & 0.937 & 0.825 & 0.128\\  \hline
     Depth DRIUnet & \textbf{0.068} & 0.964 & \textbf{0.974} & 0.276 & 0.936 & 0.816 &  0.138\\ \hline
     PD DRIUnet & 0.071 & \textbf{0.972} & 0.963 & \textbf{0.240} & \textbf{0.941} & \textbf{0.848} &  \textbf{0.1045}\\ \hline
     
 \end{tabular}
 %   \vspace{-2em}
  \end{table*}
%------------------------------
%-------
%Randomly selected images and the corresponding optic disc and cup delineations for different methods, along with the ground truth delineations are shown in the Fig. \ref{Fig. Origa} (for ORIGA dataset)  and Fig. \ref{Fig. Rimone} (for RIMONE and DRISHTI datasets), for qualitative comparison. 
The quantitative results obtained are tabulated in Table.II. for ORIGA dataset, including performance reported in $M-net with PT$ \cite{PT2018}, a very recent state of the art method, for comparison. We see that all our proposed networks have significantly lesser overlap error for both the cases of optic disc and cup segmentation, when compared to the other state of the art techniques. In terms of scaled accuracy, $M-net with PT$ \cite{PT2018} gives slightly better results for OD segmentation and gives similar results for our best performing network for OC segmentation. The residual blocks and DRI blocks give nearly similar results for OD segmentation but DRI blocks give much better results for OC segmentation. Also, the use of depth has no effect or in some cases degrades the performance of OD segmentation, but is seen to improve the performance of OC segmentation. This seems natural because of the depth discontinuity in the retinal surface while traversing from disc region to the cup region. Also, the CDR error is lower than (in Depth ResUnet case, equal to) the work presented in \cite{PT2018}. Since CDR happens to be one of the important parameters, the proposed networks are promising for improved glaucoma detection. Also, the use of depth based CRF or image based CRF or even combined CRF for post-processing always seemed to give inferior results. A reason for this could be lack of marked transitions in either intensities or depth of retinal images. This shows that multimodal fusion networks are indeed helpful in fusing information from multiple domains rather than just using CRF based post-processing. Finally, as an additional experiment to evaluate the performance of the multimodal fusion network with a different guide image, we employed pseudo-depth image as the guide in place of the depth map. Surprisingly, this gives much better performance when compared to all other experiments. The quantitative results for RIMONE and ORIGA datasets are tabulated in Table. III. Again, the best performing model seems to be the pseudo-depth model.     

% %------------------Table 3 AUC ROC
% \begin{table}
% \caption{Area Under the Curve for ROC for the three datasets }
%   \centering
% %  \renewcommand{\arraystretch}{1.2}
%   \begin{tabular}{|c|c|c|c|}
%     \hline
%     \textbf{Method}& 
%     \textbf{ORIGA} & \textbf{RIMONE} \\
%     \hline 
%     % \textbf{Inactive Modes} & \textbf{Description}\\
%      ResUnet & 0.8139 & 0.8816   \\ \hline
%      DRIUnet & 0.8167 & 0.9246   \\ \hline
%      DRIUnet CRF & 0.8188 & 0.9219   \\ \hline
%      Depth ResUnet & 0.8139 & 0.9039\\  \hline
%      Depth DRIUnet & 0.8225 & 0.8868   \\ \hline
%      PD DRIUnet & 0.8404 & 0.9390\\ \hline
%  \end{tabular}
%  %   \vspace{-2em}
%   \end{table}
Additionally, we also performed a binary classification of whether the given retinal image is glaucomatous or not, by a simple thresholding operation on vertical CDR. Retinal images with vertical CDR greater than $0.6$  is usually classified as glaucomatous. We do the experimentation on the test set of two datasets - ORIGA and RIMONE. The Drishti dataset lacks sufficient number of images for testing and moreover lacks the ground truth data for glaucoma classification. The receiver operating characteristic (ROC) curves are shown in Figure. \ref{Fig. ROC} for the two datasets and for different experiments. For ORIGA dataset, the lowest achieved area under the curve (AUC) score is $0.8139$ for the ResUnet network and for both the cases with depth and without depth. The DRIUnet performs better than ResUnet and both CRF and depth guidance improved the DRIUnet performance for glaucoma classification in terms of AUC. The best performing model is the pseudo-depth model with AUC of $0.8404$. The best performing model of the work in \cite{PT2018} achieves an AUC of $0.8508$. We believe that our models can achieve similar results simply by increasing the batch size during training (in this work our batch size was only $10$ considering the system limitations). Further the authors in \cite{PT2018} use the segmentation outputs to train again for glaucoma classification, while we choose to perform simple thresholding of CDR for glaucoma classification.

\section{Conclusion}
In this work, we proposed for the first time, a deep learning framework for monocular retinal depth estimation. We proposed a fully convolutional network (FCN) architecture for depth estimation and also a novel block called dilated residual inception (DRI) block for simultaneous multiscale feature extraction. To overcome the problem of limited data for depth estimation, we also proposed a new pretraining scheme called pseudo-depth reconstruction specifically for the depth estimation task. The proposed pretraining scheme is empirically shown to give superior results when compared with standard pretraining technique using a denoising autoencoder. 

Subsequently, we proposed a multimodal fusion block to extract and fuse features from two different modalities. Then we proposed a fully convolutional guided network that utilizes multimodal fusion block for the task of semantic segmentation. In our case, we used the guided network for optic disc (OD) and optic cup (OC) segmentation for color fundus image with depth map as the guide. The proposed framework can be utilized for any semantic segmentation task and with any two different modalities. Finally, we evaluated the suitability of pseudo-depth image as guide for OD-OC segmentation and it was shown to give superior results compared to using the depth map as the guide, with the depth being estimated from the proposed FCN. This was verified with experiments on three standard datasets. 

In future, it would be interesting to further explore other self-supervised learning techniques for representation learning, given the lack of availability of large number of images in the medical imaging domain. Also, availability of a larger dataset for depth estimation and also a dataset with ground truth data for both depth and OD-OC segmentation would help us conclude more concretely on suitability of depth map for OD-OC segmentation task in a FCN framework. Our proposed multimodal network could be extended to use any set of input images that has complimentary information.
Specifically, it would be interesting to explore other image modalities (such as OCT used in \cite{depthmultimodal}) with our FCN framework for segmentation.

% or


% \section*{Acknowledgment}


\ifCLASSOPTIONcaptionsoff
  \newpage
\fi



 %\newpage
\bibliographystyle{IEEEtran}
%\begin{thebibliography}{1}
%\section{REFERENCES}
\label{sec:ref}
------
\bibliography{refs}

\begin{thebibliography}{10}
\providecommand{\url}[1]{#1}
\csname url@samestyle\endcsname
\providecommand{\newblock}{\relax}
\providecommand{\bibinfo}[2]{#2}
\providecommand{\BIBentrySTDinterwordspacing}{\spaceskip=0pt\relax}
\providecommand{\BIBentryALTinterwordstretchfactor}{4}
\providecommand{\BIBentryALTinterwordspacing}{\spaceskip=\fontdimen2\font plus
\BIBentryALTinterwordstretchfactor\fontdimen3\font minus
  \fontdimen4\font\relax}
\providecommand{\BIBforeignlanguage}[2]{{%
\expandafter\ifx\csname l@#1\endcsname\relax
\typeout{** WARNING: IEEEtran.bst: No hyphenation pattern has been}%
\typeout{** loaded for the language `#1'. Using the pattern for}%
\typeout{** the default language instead.}%
\else
\language=\csname l@#1\endcsname
\fi
#2}}
\providecommand{\BIBdecl}{\relax}
\BIBdecl

\bibitem{Clinicalref1}
P.~Hrynchak, N.~Hutchings, D.~Jones, and T.~Simpson, ``A comparison of
  cup-to-disc ratio measurement in normal subjects using optical coherence
  tomography image analysis of the optic nerve head and stereo fundus
  biomicroscopy,'' \emph{Ophthalmic and Physiological Optics}, vol.~24, no.~6,
  pp. 543--550, 2004.

\bibitem{Clinicalref2}
J.~Xu, H.~Ishikawa, G.~Wollstein, R.~A. Bilonick, K.~R. Sung, L.~Kagemann,
  K.~A. Townsend, and J.~S. Schuman, ``Automated assessment of the optic nerve
  head on stereo disc photographs,'' \emph{Investigative ophthalmology \&
  visual science}, vol.~49, no.~6, pp. 2512--2517, 2008.

\bibitem{PAMIstereoref}
L.~Tang, M.~K. Garvin, K.~Lee, W.~L. Alward, Y.~H. Kwon, and M.~D. Abramoff,
  ``Robust multiscale stereo matching from fundus images with radiometric
  differences,'' \emph{IEEE transactions on pattern analysis and machine
  intelligence}, vol.~33, no.~11, pp. 2245--2258, 2011.

\bibitem{origa}
Z.~Zhang, F.~S. Yin, J.~Liu, W.~K. Wong, N.~M. Tan, B.~H. Lee, J.~Cheng, and
  T.~Y. Wong, ``Origa-light: An online retinal fundus image database for
  glaucoma analysis and research,'' in \emph{Engineering in Medicine and
  Biology Society (EMBC), 2010 Annual International Conference of the
  IEEE}.\hskip 1em plus 0.5em minus 0.4em\relax IEEE, 2010, pp. 3065--3068.

\bibitem{rimone}
F.~Fumero, S.~Alay{\'o}n, J.~Sanchez, J.~Sigut, and M.~Gonzalez-Hernandez,
  ``Rim-one: An open retinal image database for optic nerve evaluation,'' in
  \emph{Computer-Based Medical Systems (CBMS), 2011 24th International
  Symposium on}.\hskip 1em plus 0.5em minus 0.4em\relax IEEE, 2011, pp. 1--6.

\bibitem{drishti}
J.~Sivaswamy, S.~Krishnadas, G.~D. Joshi, M.~Jain, and A.~U.~S. Tabish,
  ``Drishti-gs: Retinal image dataset for optic nerve head (onh)
  segmentation,'' in \emph{Biomedical Imaging (ISBI), 2014 IEEE 11th
  International Symposium on}.\hskip 1em plus 0.5em minus 0.4em\relax IEEE,
  2014, pp. 53--56.

\bibitem{temp-matching}
A.~Aquino, M.~E. Geg{\'u}ndez-Arias, and D.~Mar{\'\i}n, ``Detecting the optic
  disc boundary in digital fundus images using morphological, edge detection,
  and feature extraction techniques,'' \emph{IEEE transactions on medical
  imaging}, vol.~29, no.~11, pp. 1860--1869, 2010.

\bibitem{snakes}
J.~Lowell, A.~Hunter, D.~Steel, A.~Basu, R.~Ryder, E.~Fletcher, and L.~Kennedy,
  ``Optic nerve head segmentation,'' \emph{IEEE Transactions on medical
  Imaging}, vol.~23, no.~2, pp. 256--264, 2004.

\bibitem{levelsets}
D.~Wong, J.~Liu, J.~Lim, X.~Jia, F.~Yin, H.~Li, and T.~Wong, ``Level-set based
  automatic cup-to-disc ratio determination using retinal fundus images in
  argali,'' in \emph{Engineering in Medicine and Biology Society, 2008. EMBS
  2008. 30th Annual International Conference of the IEEE}.\hskip 1em plus 0.5em
  minus 0.4em\relax IEEE, 2008, pp. 2266--2269.

\bibitem{superpixel}
J.~Cheng, J.~Liu, Y.~Xu, F.~Yin, D.~W.~K. Wong, N.-M. Tan, D.~Tao, C.-Y. Cheng,
  T.~Aung, and T.~Y. Wong, ``Superpixel classification based optic disc and
  optic cup segmentation for glaucoma screening,'' \emph{IEEE Transactions on
  Medical Imaging}, vol.~32, no.~6, pp. 1019--1032, 2013.

\bibitem{Rbend}
G.~D. Joshi, J.~Sivaswamy, and S.~Krishnadas, ``Optic disk and cup segmentation
  from monocular color retinal images for glaucoma assessment,'' \emph{IEEE
  transactions on medical imaging}, vol.~30, no.~6, pp. 1192--1205, 2011.

\bibitem{DepthDictionary}
A.~Chakravarty and J.~Sivaswamy, ``Coupled sparse dictionary for depth-based
  cup segmentation from single color fundus image,'' in \emph{International
  Conference on Medical Image Computing and Computer-Assisted
  Intervention}.\hskip 1em plus 0.5em minus 0.4em\relax Springer, 2014, pp.
  747--754.

\bibitem{DepthAkshaya}
A.~Ramaswamy, K.~Ram, and M.~Sivaprakasam, ``A depth based approach to glaucoma
  detection using retinal fundus images,'' 2016.

\bibitem{depth1}
D.~Eigen, C.~Puhrsch, and R.~Fergus, ``Depth map prediction from a single image
  using a multi-scale deep network,'' in \emph{Advances in neural information
  processing systems}, 2014, pp. 2366--2374.

\bibitem{depth2}
F.~Liu, C.~Shen, and G.~Lin, ``Deep convolutional neural fields for depth
  estimation from a single image,'' in \emph{Proceedings of the IEEE Conference
  on Computer Vision and Pattern Recognition}, 2015, pp. 5162--5170.

\bibitem{DepthHuber1}
I.~Laina, C.~Rupprecht, V.~Belagiannis, F.~Tombari, and N.~Navab, ``Deeper
  depth prediction with fully convolutional residual networks,'' in \emph{3D
  Vision (3DV), 2016 Fourth International Conference on}.\hskip 1em plus 0.5em
  minus 0.4em\relax IEEE, 2016, pp. 239--248.

\bibitem{depthmultimodal}
M.~S. Miri, M.~D. Abr{\`a}moff, K.~Lee, M.~Niemeijer, J.-K. Wang, Y.~H. Kwon,
  and M.~K. Garvin, ``Multimodal segmentation of optic disc and cup from sd-oct
  and color fundus photographs using a machine-learning graph-based approach,''
  \emph{IEEE transactions on medical imaging}, vol.~34, no.~9, pp. 1854--1866,
  2015.

\bibitem{zilly2017glaucoma}
J.~Zilly, J.~M. Buhmann, and D.~Mahapatra, ``Glaucoma detection using entropy
  sampling and ensemble learning for automatic optic cup and disc
  segmentation,'' \emph{Computerized Medical Imaging and Graphics}, vol.~55,
  pp. 28--41, 2017.

\bibitem{Resunet}
S.~M. Shankaranarayana, K.~Ram, K.~Mitra, and M.~Sivaprakasam, ``Joint optic
  disc and cup segmentation using fully convolutional and adversarial
  networks,'' in \emph{Fetal, Infant and Ophthalmic Medical Image
  Analysis}.\hskip 1em plus 0.5em minus 0.4em\relax Springer, 2017, pp.
  168--176.

\bibitem{PT2018}
H.~Fu, J.~Cheng, Y.~Xu, D.~W.~K. Wong, J.~Liu, and X.~Cao, ``Joint optic disc
  and cup segmentation based on multi-label deep network and polar
  transformation,'' \emph{IEEE Transactions on Medical Imaging}, 2018.

\bibitem{unet}
O.~Ronneberger, P.~Fischer, and T.~Brox, ``U-net: Convolutional networks for
  biomedical image segmentation,'' in \emph{International Conference on Medical
  image computing and computer-assisted intervention}.\hskip 1em plus 0.5em
  minus 0.4em\relax Springer, 2015, pp. 234--241.

\bibitem{Autoencoder1}
P.~Vincent, H.~Larochelle, Y.~Bengio, and P.-A. Manzagol, ``Extracting and
  composing robust features with denoising autoencoders,'' in \emph{Proceedings
  of the 25th international conference on Machine learning}.\hskip 1em plus
  0.5em minus 0.4em\relax ACM, 2008, pp. 1096--1103.

\bibitem{ContextAE}
D.~Pathak, P.~Krahenbuhl, J.~Donahue, T.~Darrell, and A.~A. Efros, ``Context
  encoders: Feature learning by inpainting,'' in \emph{Proceedings of the IEEE
  Conference on Computer Vision and Pattern Recognition}, 2016, pp. 2536--2544.

\bibitem{Colorization}
G.~Larsson, M.~Maire, and G.~Shakhnarovich, ``Colorization as a proxy task for
  visual understanding,'' in \emph{CVPR}, vol.~2, 2017, p.~8.

\bibitem{resnet}
K.~He, X.~Zhang, S.~Ren, and J.~Sun, ``Deep residual learning for image
  recognition,'' in \emph{Proceedings of the IEEE conference on computer vision
  and pattern recognition}, 2016, pp. 770--778.

\bibitem{inception}
C.~Szegedy, S.~Ioffe, V.~Vanhoucke, and A.~A. Alemi, ``Inception-v4,
  inception-resnet and the impact of residual connections on learning.'' in
  \emph{AAAI}, 2017, pp. 4278--4284.

\bibitem{DilatedMain}
F.~Yu and V.~Koltun, ``Multi-scale context aggregation by dilated
  convolutions,'' \emph{arXiv preprint arXiv:1511.07122}, 2015.

\bibitem{Atrous1}
L.-C. Chen, G.~Papandreou, I.~Kokkinos, K.~Murphy, and A.~L. Yuille, ``Deeplab:
  Semantic image segmentation with deep convolutional nets, atrous convolution,
  and fully connected crfs,'' \emph{arXiv preprint arXiv:1606.00915}, 2016.

\bibitem{Atrous2}
L.-C. Chen, G.~Papandreou, F.~Schroff, and H.~Adam, ``Rethinking atrous
  convolution for semantic image segmentation,'' \emph{arXiv preprint
  arXiv:1706.05587}, 2017.

\bibitem{BerHu1}
A.~B. Owen, ``A robust hybrid of lasso and ridge regression,''
  \emph{Contemporary Mathematics}, vol. 443, no.~7, pp. 59--72, 2007.

\bibitem{Fusenet}
C.~Hazirbas, L.~Ma, C.~Domokos, and D.~Cremers, ``Fusenet: Incorporating depth
  into semantic segmentation via fusion-based cnn architecture,'' in
  \emph{Asian Conference on Computer Vision}.\hskip 1em plus 0.5em minus
  0.4em\relax Springer, 2016, pp. 213--228.

\bibitem{Potts}
P.~Kohli, M.~P. Kumar, and P.~H. Torr, ``P3 \& beyond: Solving energies with
  higher order cliques,'' in \emph{Computer Vision and Pattern Recognition,
  2007. CVPR'07. IEEE Conference on}.\hskip 1em plus 0.5em minus 0.4em\relax
  IEEE, 2007, pp. 1--8.

\bibitem{CRF}
P.~Kr{\"a}henb{\"u}hl and V.~Koltun, ``Efficient inference in fully connected
  crfs with gaussian edge potentials,'' in \emph{Advances in neural information
  processing systems}, 2011, pp. 109--117.

\bibitem{drions}
E.~J. Carmona, M.~Rinc{\'o}n, J.~Garc{\'\i}a-Feijo{\'o}, and J.~M.
  Mart{\'\i}nez-de-la Casa, ``Identification of the optic nerve head with
  genetic algorithms,'' \emph{Artificial Intelligence in Medicine}, vol.~43,
  no.~3, pp. 243--259, 2008.

\end{thebibliography}
%\end{thebibliography}


%\end{IEEEbiography}


%\newpage



%\vfill


\end{document}